\documentclass[aps,pra,noshowpacs,twocolumn,preprintnumbers,superscriptaddress,floatfix]{revtex4-2}
\usepackage{bm}
\usepackage{amsmath, amssymb, amsthm}
\usepackage{graphicx}
\usepackage[colorlinks=true, citecolor=blue, urlcolor=blue, linkcolor=blue]{hyperref}
\usepackage{braket}

\begin{document}
\title{Understanding interaction-driven transport in flux lattices with evolution-path symmetry}

\author{Jian-Song Pan}
\email{panjsong@scu.edu.cn}
\affiliation{College of Physics, Sichuan University, Chengdu 610065, China}
\affiliation{Key Laboratory of High Energy Density Physics and Technology of Ministry of Education, Sichuan University, Chengdu 610065, China}

\author{Xiaofan Zhou}
\email{zhouxiaofan@sxu.edu.cn}
\affiliation{State Key Laboratory of Quantum Optics and Quantum Optics Devices, Institute of Laser Spectroscopy, Shanxi University, Taiyuan 030006, China}
\affiliation{Collaborative Innovation Center of Extreme Optics, Shanxi University, Taiyuan, Shanxi 030006, China}

\author{Wei Yi}
\affiliation{CAS Key Laboratory of Quantum Information, University of Science and Technology of China, Hefei 230026, China}
\affiliation{Anhui Province Key Laboratory of Quantum Network, University of Science and Technology of China, Hefei 230026, China}
\affiliation{CAS Center For Excellence in Quantum Information and Quantum Physics, Hefei 230026, China}
\affiliation{Hefei National Laboratory, University of Science and Technology of China, Hefei 230088, China}

\begin{abstract}
The destruction of Aharonov-Bohm (AB) caging by interaction and the emergence of interaction-induced chiral currents in flux lattices are two paradigmatic examples of interaction-driven quantum transport. While various mechanisms, such as bound-state formation and chiral spectral imbalance, have been proposed, a unifying physical picture remains elusive. Here, we employ the concept of \textit{evolution-path symmetry} (EPS) and its interaction-induced breaking as a framework to understand interaction-induced delocalization in flux lattices. EPS is defined as the invariance of a path's contribution under combined geometric and phase transformations. We demonstrate that in a $\pi$-flux rhombic lattice, interactions break the EPS present in the non-interacting limit by modifying the phase accumulation of many-body paths, thereby lifting the destructive interference responsible for AB caging. Furthermore, we apply this framework to explain interaction-induced chiral transport in flux ladders, where interactions break the phase relationship between symmetric paths, leading to a non-vanishing chiral current. Our work establishes EPS as a powerful tool for understanding transport phenomena beyond conventional eigenstate analysis.
\end{abstract}

\maketitle

\section{Introduction}
Localization phenomena can arise in various systems through distinct mechanisms. Anderson localization, as a typical example of localization, arises from the destructive interference among multiple scattering paths induced by disorder.~\cite{anderson1958absence,brandes2003the,Wiersma1997, Storzer2006, Scheffold1999, Schwartz2007, Lahini2008, Karbasi2012, Karbasi2014, Billy2008, Roati2008, Ludlam2005, Conti2008, Hu2008, Chabe2008, Ying2016, Choi2018, Skipetrov2016}. In clean systems, localization can also be enforced by geometry and gauge fields, as exemplified by the Aharonov-Bohm (AB) cage effect in lattices with $\pi$ flux per plaquette~\cite{Aharonov1959, vidal1998aharonov,Abilio1999,Naud2001,Li2022,Li2025}. Here, destructive interference between paths encircling a plaquette leads to perfectly flat bands and complete single-particle localization. Despite their different origins, both mechanisms rely fundamentally on wave interference. The introduction of interactions fundamentally alters this landscape. Although disordered systems may still fail to reach thermal equilibrium in the presence of interactions~\cite{Nandkishore2015,Fleishman1980,Altshuler1997,Gornyi2005,Basko2006,Oganesyan2007,Pal2010,Imbrie2016}, for a AB-caged systems with flat-band localization, inter-particle interaction can hybridize degenerate flat-band states, generate dispersion, and thereby enable transport, breaking the geometric confinement~\cite{vidal2000interaction, vidal2001disorder,Tovmasyan2013,Tovmasyan2018, cartwright2018rhombi, liberto2019nonlinear, danieli2021quantum, danieli2021nonlinear}.

The prevailing understanding for interaction-induced delocalization in AB-caged systems invokes the formation of bound states, such as doublons. A bound pair of particles, when treated as a composite object, experiences an effective magnetic flux that is the sum of the individual particle fluxes. For a $\pi$-flux lattice, a tightly-bound doublon thus senses a $2\pi$ flux, nullifying the Aharonov-Bohm phase and escaping the cage \cite{vidal2000interaction,tovmasyan2018preformed}. This picture has been corroborated by experiments in superconducting circuits \cite{martinez2023flat} and with Rydberg atoms \cite{chen2025interaction} recently.

The situation becomes more complex in other flux-lattice geometries. In flux ladders (quasi-1D counterparts of the Harper-Hofstadter model), interactions can induce chiral currents even from localized initial states \cite{tai2017microscopy}. The mechanism here is often attributed to an interaction-induced redistribution of spectral weight onto chiral eigenstates \cite{tai2017microscopy,pan2025reversal}. In the strong-interaction regime, interactions can also drive the system ground state into vortex crystal phases, where the spontaneous formation of charge density waves effectively doubles the magnetic flux sensed by each vortex, leading to phenomena such as chiral current reversal \cite{greschner2015spontaneous,kolley2015strongly,uchino2016analytical}. These diverse explanations--bound-state flux summation, spectral weight redistribution, and vortex-crystal-induced flux doubling--highlight the need for a unifying dynamical principle.

In this work, we propose the concept of \textit{evolution-path symmetry} (EPS) and its interaction-induced breaking as a unifying framework for understanding interaction-driven transport of a local initial state in flux lattices. By analyzing quantum dynamics at the level of interfering trajectories in Fock space, we demonstrate that interactions break specific symmetries in the ensemble of Feynman evolution paths, which are responsible for localization or the suppression of chiral transport in the non-interacting limit. We first illustrate this mechanism in the canonical $\pi$-flux rhombic lattice, showing how interactions selectively modify the phase accumulation along paths involving double occupancy, thereby lifting the destructive interference underlying the Aharonov-Bohm cage. We then generalize this observation to formulate the EPS concept and classify different types of symmetries. Finally, we apply the same framework to a flux ladder, explaining how interactions break the phase symmetry between pairs of paths related by chiral reflection, leading to the emergence of a finite chiral current.

The remainder of the paper is organized as follows. In Sec.~\ref{sec:rhombic}, we introduce the $\pi$-flux rhombic lattice model and demonstrate interaction-induced delocalization beyond the Aharonov-Bohm cage. The concept of evolution-path symmetry (EPS) and its classification are then developed in Sec.~\ref{sec:eps}. In Sec.~\ref{sec:many-body EPS}, we extend the concept of EPS into the many-body case. In Sec.~\ref{sec:ladder}, we apply the EPS framework to explain interaction-induced chiral transport in a flux ladder, highlighting the breaking of conjugate-phase symmetry. Finally, we summarize our findings and discuss the broader implications of the EPS approach in Sec.~\ref{sec:conclusion}.

\section{Interaction-induced delocalization in the $\pi$-flux rhombic lattice}\label{sec:rhombic}

\begin{figure}[t]
\centering
\includegraphics[width=0.9\linewidth]{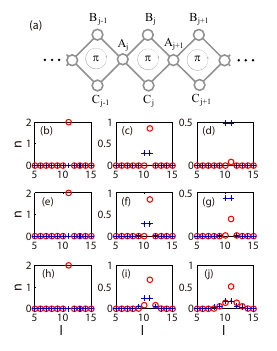}
\caption{(a) Schematic of the $\pi$-flux rhombic lattice. Arrows indicate the phase of hopping amplitudes: $e^{\pm i\pi/4}$. (b)-(d): Time evolution of the particle density at different moments ($t=0$, $1$, and $2$) for an initial doublon at a central A site, with $U=0$. Here the site labels are expanded into a series of integers: for example, $A_1$, $B_1$, and $C_1$ are labeled with $l=1$, 2, and 3, and so on. The red circles and blue crosses represent the real and imaginary parts of the wave functions. (e)-(g) Same as (b)-(d) but with $U=0.5$. (h)-(j) Density distribution for $U=1.0$. Interactions clearly facilitate delocalization beyond the AB cage. {Here we set $J=1$ as the unit of energy. The system size is $L=20$ (i.e., $3L=60$ sites), and periodic boundary conditions are employed. The calculations are performed using exact diagonalization.}}
\label{fig:rhombic}
\end{figure}

We will begin with an example of interaction-induced transport in a rhombic lattice. This example provides initial insight into the concept of evolution-path symmetry and its breaking due to interactions. We will then generalize the concept in section IV and finally illustrate it with a more complex example where interaction-induced chiral transport is observed. Although these phenomena are generally explained by the microscopic properties of the model, we aim to interpret them within a unified framework with the concept of EPS.

We consider a rhombic lattice with a $\pi$ flux per plaquette, which exhibits Aharonov-Bohm caging, as shown in Fig.~\ref{fig:rhombic}(a). Under a symmetric gauge, the tight-binding Hamiltonian for spinless bosons is $H=H_{0}+H_{\text{int}}$ with
\begin{align}
H_0 &= J e^{i\pi/4} \sum_j \bigl( \hat{c}_{j,B}^\dagger \hat{c}_{j,A}
      + \hat{c}_{j,A}^\dagger \hat{c}_{j,C} + \hat{c}_{j+1,A}^\dagger \hat{c}_{j,B} \nonumber \\
    &\quad + \hat{c}_{j,C}^\dagger \hat{c}_{j+1,A} \bigr) + \text{H.c.}, \label{eq:H0_rhombic} \\
H_{\text{int}} &= \frac{U}{2} \sum_{j,\xi=A,B,C} \hat{n}_{j,\xi} (\hat{n}_{j,\xi} - 1), \label{eq:Hint_rhombic}
\end{align}
where $\hat{c}_{j,\xi}$ annihilates a boson at site $\xi$ of unit cell $j$, and $\hat{n}_{j,\xi} = \hat{c}_{j,\xi}^\dagger \hat{c}_{j,\xi}$. The single-particle spectrum consists of a perfectly flat band at energy $E=0$ and two dispersive bands, leading to the AB cage effect: any localized single-particle wavepacket remains dynamically confined.

Figs.~\ref{fig:rhombic}(b)-(j) show the time evolution of a two-particle doublon initially placed on a single A site. In the non-interacting case ($U=0$), the doublon exhibits only minimal spreading, consistent with the AB cage (though a careful re-examination of our earlier numerics revealed that any apparent delocalization was due to finite-size and boundary effects; true asymptotic localization holds). When interactions are turned on ($U=0.5, 1.0$), the doublon rapidly delocalizes across the lattice. This confirms that interactions are necessary to break the AB cage for a compact initial state, in agreement with the bound-state picture \cite{vidal2000interaction,martinez2023flat}.

\begin{figure}[t]
\centering
\includegraphics[width=0.95\linewidth]{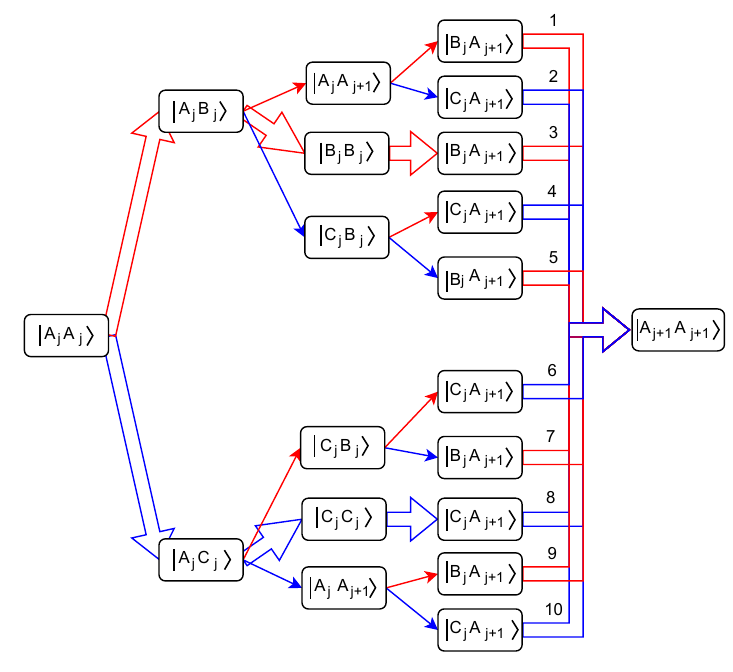}
\caption{Shortest evolution paths in two-particle Fock space connecting the doublon state $\ket{A_j A_j}$ to $\ket{A_{j+1} A_{j+1}}$ on the rhombic lattice. Each path consists of four hopping steps. Thin arrows represent single-particle hopping with amplitude $1$; bold arrows represent correlated hopping with amplitude $\sqrt{2}$ due to bosonic enhancement. Red (blue) arrows indicate a phase factor of $e^{i\pi/4}$ ($e^{-i\pi/4}$). The total phase factor for each path (from paths 1 to 10) is $-2, 2, -4, 2, 2, 2, 2, -4, 2, -2$ (i.e., the products of the hopping coefficients on the paths by setting $J=1$), respectively. The summation over all paths gives zero, indicating destructive interference in the non-interacting limit. Paths 3 and 6 (highlighted) involve intermediate double occupancy and thus acquire additional interaction-induced phase shifts when $U \neq 0$.}
\label{fig:paths}
\end{figure}

To understand the mechanism underlying interaction-induced delocalization, we analyze the shortest evolution paths connecting neighboring doublon states, which are assumed to dominate the evolution in a short enough time. Fig.~\ref{fig:paths} enumerates the ten shortest paths connecting the doublon state $\ket{A_j A_j}$ to $\ket{A_{j+1} A_{j+1}}$ in the rhombic lattice. In the non-interacting limit, the total phase factor summed over all ten paths is exactly zero, indicating complete destructive interference. This explains why the doublon remains localized in the absence of interactions, as observed in Fig.~\ref{fig:rhombic}(b)-(d).

The key effect of interactions is to introduce an on-site phase factor $e^{-i U n (n-1) \Delta t / 2}$ for a site with occupation $n$. This phase depends on the occupation history along a path. In the two-particle case, only paths that involve intermediate double occupancy acquire such interaction-induced phase shifts. Specifically, paths 3 and 6 in Fig.~\ref{fig:paths} pass through configurations where both particles occupy the same site (double occupancy). When $U \neq 0$, these paths accumulate an additional phase factor $e^{-i U \Delta t}$ (for $n=2$), while all other paths (without double occupancy) remain unaffected.

Consequently, the delicate balance of phase factors that enforces destructive interference in the non-interacting limit is broken. The total transition amplitude between doublon states $\ket{A_j A_j}$ and $\ket{A_{j+1} A_{j+1}}$ becomes finite, enabling delocalization. This mechanism explains why interactions are necessary to break the AB cage for a compact doublon initial state, and why the delocalization rate increases with interaction strength $U$, as observed in Fig.~\ref{fig:rhombic}(e)-(j).

The path analysis reveals a fundamental symmetry breaking: in the non-interacting case, the paths are organized into symmetry-related pairs that cancel each other's contributions. Interactions break this symmetry by selectively modifying the phase accumulation along specific paths that involve double occupancy. This observation motivates us to formalize the concept of EPS.

\section{Evolution-path symmetry: Concept and classification}\label{sec:eps}

The path interference analysis of the $\pi$-flux rhombic lattice suggests a general principle: localization and transport suppression in non-interacting systems often result from symmetries among evolution paths. To systematize this idea, we formulate the notion of EPS within the path-integral framework.

While the time evolution of a quantum state is typically expressed via the time evolution operator~\cite{Wang2008},
\begin{equation}\label{eq:time_evolution}
|\psi(t)\rangle = U(t,t_0) |\psi(0)\rangle = \mathcal{T} \exp\left(-i \int_{t_0}^{t} d\tau \, H(\tau)/\hbar \right) |\psi(0)\rangle,
\end{equation}
where $\mathcal{T}$ is the time ordering operator, we focus instead on the symmetries emergent in the evolution path integral representation.

For simplicity, consider first a single particle on a lattice. The propagator from an initial site $\boldsymbol{i}$ to a final site $\boldsymbol{j}$ is given by
\begin{equation}\label{eq:propagator}
K(\boldsymbol{j},t;\boldsymbol{i},t_0) = \langle \boldsymbol{j} | U(t,t_0) | \boldsymbol{i} \rangle.
\end{equation}
By slicing the time interval into $N$ segments, we can write
\begin{equation}\label{eq:propagator_split}
\begin{split}
K(\boldsymbol{j},t;\boldsymbol{i},t_0) &= \lim_{N \to \infty} \sum_{\boldsymbol{j}_1} \cdots \sum_{\boldsymbol{j}_{N-1}} \langle \boldsymbol{j} | e^{-i H(t_{N-1}) \Delta t / \hbar} | \boldsymbol{j}_{N-1} \rangle \\
&\quad \times \langle \boldsymbol{j}_{N-1} | e^{-i H(t_{N-2}) \Delta t / \hbar} | \boldsymbol{j}_{N-2} \rangle \\
&\quad \times \cdots \times \langle \boldsymbol{j}_1 | e^{-i H(t_0) \Delta t / \hbar} | \boldsymbol{i} \rangle,
\end{split}
\end{equation}
where $t_n = t_0 + n \Delta t$ and $\Delta t = (t - t_0)/N$. Each term in the sum corresponds to a particular spacetime path, and the total amplitude is a coherent sum over all such evolution paths, each weighted by a phase factor $\Phi[\mathcal{P}]$.

\begin{figure*}[t]
\centering
\includegraphics[width=1\linewidth]{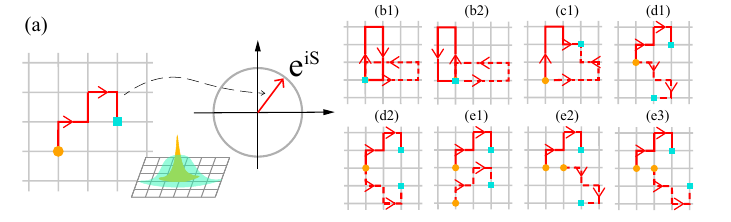}
\caption{\textbf{Examples of evolution-path symmetries in a square lattice.}
(a) General illustration of evolution paths and their phase factors, which govern the dynamical evolution.
(b1) Paths with the same initial and final site, related by reflection symmetry.
(b2) Paths with the same initial and final site, related by rotation symmetry.
(c1) Paths with the same initial and final sites (but different sites), related by reflection symmetry.
(d1) Paths with the same initial site but different final sites, related by rotation symmetry.
(d2) Paths with the same initial site but different final sites, related by reflection symmetry.
(e1) Paths with different initial and final sites, related by translation symmetry.
(e2) Paths with different initial and final sites, related by combined translation and rotation symmetry.
(e3) Paths with different initial and final sites, related by combined translation and reflection symmetry.
These geometric relations, together with the phase relationship between symmetric paths (which is determined by the system Hamiltonian), give rise to the definition of EPS.}
\label{fig:path_pairs}
\end{figure*}

We assume that particle number is conserved, so the dynamics is confined to the single-particle Fock space. For simplicity, we only consider time-independent Hamiltonian. Writing $H = H_{\mathrm{hop}} + H_{\mathrm{on}}$, where $H_{\mathrm{hop}}$ contains inter-site hopping and $H_{\mathrm{on}}$ contains on-site terms, we have $H_{\mathrm{hop}}|\boldsymbol{j}\rangle = \sum_{\boldsymbol{j}'\neq\boldsymbol{j}} J_{\boldsymbol{j}',\boldsymbol{j}} |\boldsymbol{j}'\rangle$ and $H_{\mathrm{on}}|\boldsymbol{j}\rangle = \mu_{\boldsymbol{j}} |\boldsymbol{j}\rangle$, where { $J_{\boldsymbol{j}'\boldsymbol{j}}$ is the hopping coefficient for the hopping term $J_{\boldsymbol{j}'\boldsymbol{j}} |\boldsymbol{j}'\rangle\langle \boldsymbol{j}|$}. Expanding $e^{-i H(t_{j})\Delta t/\hbar} \approx 1 - iH(t_{j})\Delta t/\hbar$ for small $\Delta t$ (taking $N \to \infty$), we obtain
\begin{equation}\label{eq:first_step}
\begin{split}
&\sum_{\boldsymbol{j}_{1}}|\boldsymbol{j}_{1}\rangle\langle \boldsymbol{j}_{1}|e^{-i H(t_0) \Delta t / \hbar}|\boldsymbol{i}\rangle\\
&=\left(1-\frac{i\Delta t}{\hbar}\mu_{\boldsymbol{i}}\right)|\boldsymbol{i}\rangle-\frac{i\Delta t}{\hbar}\sum_{\boldsymbol{j}_{1}\neq \boldsymbol{i}}J_{\boldsymbol{j}_{1},\boldsymbol{i}}|\boldsymbol{j}_{1}\rangle\\
&=|\boldsymbol{i}\rangle-\frac{i\Delta t}{\hbar}\sum_{\boldsymbol{j}_{1}}J_{\boldsymbol{j}_{1},\boldsymbol{i}}|\boldsymbol{j}_{1}\rangle,
\end{split}
\end{equation}
where we define $J_{\boldsymbol{l},\boldsymbol{l}}=\mu_{\boldsymbol{l}}$. Furthermore,
\begin{equation}\label{eq:second_step}
\begin{split}
&\sum_{\boldsymbol{j}_{1}}\sum_{\boldsymbol{j}_{2}}|\boldsymbol{j}_{2}\rangle\langle \boldsymbol{j}_{1}|e^{-i H(t_0) \Delta t / \hbar}|\boldsymbol{j}_{1}\rangle\langle \boldsymbol{j}_{1}|e^{-i H(t_0) \Delta t / \hbar}|\boldsymbol{i}\rangle\\
&=|\boldsymbol{i}\rangle-\frac{i\Delta t}{\hbar}\sum_{\boldsymbol{j}_{1}}J_{\boldsymbol{j}_{1},\boldsymbol{i}}|\boldsymbol{j}_{1}\rangle-\frac{i\Delta t}{\hbar}\sum_{\boldsymbol{j}_{2}}J_{\boldsymbol{j}_{2},\boldsymbol{i}}|\boldsymbol{j}_{2}\rangle\\
&\quad+\left(-\frac{i\Delta t}{\hbar}\right)^{2}\sum_{\boldsymbol{j}_{1}}\sum_{\boldsymbol{j}_{2}}J_{\boldsymbol{j}_{2},\boldsymbol{j}_{1}}J_{\boldsymbol{j}_{1},\boldsymbol{i}}|\boldsymbol{j}_{2}\rangle.
\end{split}
\end{equation}

The propagator can be expressed as a series $K(\boldsymbol{j},t;\boldsymbol{i},t_0)=\sum_{n=0}^{N}K_{n}(\boldsymbol{j},t;\boldsymbol{i},t_0)$, where
\begin{equation}\label{eq:K_series}
\begin{split}
&K_{0}=\delta_{\boldsymbol{i},\boldsymbol{j}},\\
&K_{1}=C_{N}^{1}\left(-\frac{i\Delta t}{\hbar}\right)J_{\boldsymbol{j},\boldsymbol{i}}
      =C_{N-1}^{1}\left(-\frac{i\Delta t}{\hbar}\right)\mathcal{P}_{1}(\boldsymbol{j},\boldsymbol{i}),\\
&K_{2}=C_{N}^{2}\left(-\frac{i\Delta t}{\hbar}\right)^{2}\sum_{\boldsymbol{j}_{1}}J_{\boldsymbol{j},\boldsymbol{j}_{1}}J_{\boldsymbol{j}_{1},\boldsymbol{i}}\\
     &=C_{N-1}^{2}\left(-\frac{i\Delta t}{\hbar}\right)^{2}\sum_{\xi=1}^{N_{2}}\mathcal{P}_{2}^{(\xi)}(\boldsymbol{j},\boldsymbol{i}),\\
&\qquad\qquad\vdots\\
&K_{n}=C_{N}^{n}\left(-\frac{i\Delta t}{\hbar}\right)^{n}
       \sum_{\boldsymbol{j}_{1}}\cdots\sum_{\boldsymbol{j}_{n-1}}
       J_{\boldsymbol{j},\boldsymbol{j}_{n-1}}J_{\boldsymbol{j}_{n-1},\boldsymbol{j}_{n-2}}\cdots J_{\boldsymbol{j}_{1},\boldsymbol{i}}\\
     &=C_{N}^{n}\left(-\frac{i\Delta t}{\hbar}\right)^{n}
       \sum_{\xi=1}^{N_{n}}\mathcal{P}_{n}^{(\xi)}(\boldsymbol{j},\boldsymbol{i}),\\
&\qquad\qquad\vdots\\
&K_{N}=C_{N}^{N}\left(-\frac{i\Delta t}{\hbar}\right)^{N}
       \sum_{\xi=1}^{N_{N}}\mathcal{P}_{N}^{(\xi)}(\boldsymbol{j},\boldsymbol{i}).
\end{split}
\end{equation}
Here we have introduced the phase factor for a specific path connecting site $\boldsymbol{i}$ and site $\boldsymbol{j}$ with $n$ hopping steps, $\mathcal{P}_{n}^{(\xi)}=J_{\boldsymbol{j},\boldsymbol{j}_{n-1}}J_{\boldsymbol{j}_{n-1},\boldsymbol{j}_{n-2}}\cdots J_{\boldsymbol{j}_{1},\boldsymbol{i}}$, and $N_{n}$ denotes the number of distinct $n$-step paths between the two sites. Here, $\xi$ denotes the different connected paths between the initial and final sites. Thus we obtain the path expansion
\begin{equation}\label{eq:path_expansion}
K(\boldsymbol{j},t;\boldsymbol{i},t_0)=\sum_{n=0}^{N}C_{N}^{n}\left(-\frac{i\Delta t}{\hbar}\right)^{n}
\sum_{\xi=1}^{N_{n}}\mathcal{P}_{n}^{(\xi)}(\boldsymbol{j},\boldsymbol{i}).
\end{equation}
If the shortest distance between site $\boldsymbol{i}$ and site $\boldsymbol{j}$ is $d$, then the lowest-order non-vanishing contribution to $K(\boldsymbol{j},t;\boldsymbol{i},t_0)$ is $K_{d}$, since $K_{n<d}=0$.

Let us discuss the conditions for neglecting higher-order terms in the expansion of the propagator. Since $\Delta t=(t-t_{0})/N$, in the limit $N\rightarrow \infty$ we have
\begin{equation}\label{eq:K_n_to_n_dp1}
\frac{K_{n+1}}{K_{n}}=
-\frac{i}{\hbar}\frac{t-t_{0}}{n+1}\,
\frac{\displaystyle\sum_{\xi=1}^{N_{n+1}}\mathcal{P}_{n+1}^{(\xi)}(\boldsymbol{j},\boldsymbol{i})}
       {\displaystyle\sum_{\xi=1}^{N_{n}}\mathcal{P}_{n}^{(\xi)}(\boldsymbol{j},\boldsymbol{i})},\quad N\rightarrow\infty.
\end{equation}
The total phase factor $\Gamma_{n}=\sum_{\xi=1}^{N_{n}}\mathcal{P}_{n}^{(\xi)}(\boldsymbol{j},\boldsymbol{i})$ depends on both the geometry of the (effective) lattice and the details of the model Hamiltonian; consequently, the ratio $K_{n+1}/K_{n}$ is difficult to estimate in general.

Nevertheless, in simple cases one can gain some insight from Eq.~\eqref{eq:path_expansion} via this ratio. For example, consider a homogeneous lattice with only nearest-neighbor hopping and a constant hopping amplitude $J$, i.e., $H=J\sum_{\langle \boldsymbol{i},\boldsymbol{j}\rangle}\hat{c}_{\boldsymbol{i}}^{\dagger}\hat{c}_{\boldsymbol{j}}$. Then $\Gamma_{n}=N_{n}J^{n}$, $K_{n+1}=0$ (in fact, for any odd number $m$, $K_{n+m}=0$ and the series include only the even-higher-order terms) and $K_{n+2}=\eta (n+1) N_{n}J^{n+2}$ with the coordination number $\eta$, such that
\begin{equation}\label{eq:homogenous_lattice_KoverK}
\frac{|K_{n+2}|}{|K_{n}|}=\frac{\eta J^2(t-t_{0})^{2}}{ (n+2)\hbar^{2}},\quad N\rightarrow \infty.
\end{equation}
If the evolution time is sufficiently short, the hopping amplitude sufficiently small, or the distance $d$ between the initial and final sites sufficiently large, such that $|K_{d+1}/K_d| \ll 1$, then, apart from the lowest-order contribution $K_d$, all higher-order terms are negligible. For long-term evolution or other more complex configurations, the analysis becomes difficult and, in general, higher-order terms must be considered. Hence, a systematic analysis of the symmetry among paths is desirable for understanding the dynamics.

\begin{figure}[t]
\centering
\includegraphics[width=0.95\linewidth]{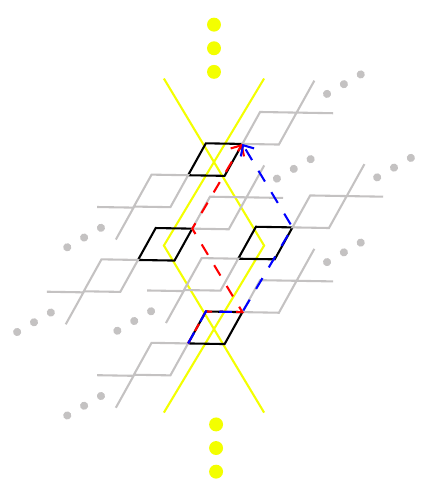}
\caption{Illustration of the product of rhombic lattices of two distinguishable particles (big yellow and small grey/black rhombic lattices). Two evolution paths denoted by red and blue dashed lines have opposite contributions, since they involve the same first half part and opposite latter half parts due to the $\pi$ flux.}
\label{fig:paths_cancellation}
\end{figure}

Further, a \textbf{EPS} exists between two paths $\mathcal{P}$ and $\mathcal{P}'$ if:
\begin{enumerate}
    \item Their real-space projections are related by a geometric symmetry belonging to the lattice symmetry group (e.g., reflection, rotation, and translation restricted by the lattice geometry);
    \item Their {amplitudes} and accumulated phase factors $\Phi[\mathcal{P}'] $ and $\Phi[\mathcal{P}]$ keep fixed relationship (e.g.,  the same or conjugate).
\end{enumerate}
The second requirement, which is mainly determined by system Hamiltonian, makes EPS a useful concept because fixed phase relations facilitate the interference between different paths. Breaking EPS, whether by local potentials, gauge fields, or interactions, can alter the system's dynamical properties.

Fig.~\ref{fig:path_pairs} illustrates some examples of EPS in a square lattice. Different lattice geometries and Hamiltonian give rise to distinct types of EPS. Two types are particularly relevant for transport phenomena:
\begin{enumerate}
    \item \textbf{Destructive-interference EPS}  $\Phi[\mathcal{P}'] =  \Phi[\mathcal{P}]\pm\pi$ {and $|\mathcal{P}'|=|\mathcal{P}|$}: Paths appear in pairs with opposite phases. Their amplitudes cancel exactly, leading to localization (e.g., AB caging in the $\pi$-flux rhombic lattice).
    \item \textbf{Unchanged-/Conjugate-phase EPS} $\Phi[\mathcal{P}'] =\pm\Phi[\mathcal{P}]$ {and $|\mathcal{P}'|=|\mathcal{P}|$}: Paths appear in pairs with unchanged or conjugate phase factors. Their contributions cancel in certain observables (e.g., chiral displacement in a flux ladder with chiral symmetry), suppressing chiral transport.
\end{enumerate}

\section{EPS on many-body Fock-state lattice}\label{sec:many-body EPS}

{The conventional lattice-symmetry-based single-particle EPS becomes inadequate in the presence of interactions. In such cases, the system evolution cannot be simply regarded as single-particle dynamics between lattice sites. The same expansion procedure elaborated in the above section remains valid for the propagator of a many-particle Fock state if the ``sites'' are taken to be many-particle Fock states. The evolution paths discussed in Sec. II correspond to the two-particle case. In other words, for the many-body case, we simply replace each single-particle lattice site with a many-particle Fock state, thereby constructing a many-body Fock-state lattice where each site corresponds to a Fock state on the original lattice. The interaction terms then become the on-site potential. }

{The EPS in the many-body Fock lattice in general originates from the single-particle symmetry. We will elaborate this point with an example with two bosons following the route: single-particle symmetry $\rightarrow$  symmetry in the Fock-state lattice of distinguishable particles $\rightarrow$ symmetry in the Fock-state lattice of indistinguishable particles. The Hilbert space of distinguishable particles is $\mathcal{H}_2^{\text{dist}} = \mathcal{H}_1 \otimes \mathcal{H}_1$, with basis $|i,j\rangle_{\text{dist}} = |i\rangle \otimes |j\rangle$. The Hamiltonian for distinguishable particles is $H_{\text{dist}} = h \otimes I + I \otimes h$, where $h$ is the single-particle Hamiltonian. The geometric symmetry $U_g$ acting on $\mathcal{H}_1$ induces a symmetry $U_g^{\text{dist}} = U_g \otimes U_g$ on $\mathcal{H}_2^{\text{dist}}$, which leaves $H_{\text{dist}}$ invariant. The symmetric (bosonic) subspace for two indistinguishable bosons is obtained by projecting onto states invariant under particle exchange:
\begin{equation}
\mathcal{H}_2^{\text{bos}} = P_{\text{sym}} \mathcal{H}_2^{\text{dist}}P_{\text{sym}},
\end{equation}
with the projection operator $P_{\text{sym}} = \frac{1}{2}(I + \text{SWAP})$, where $\text{SWAP}|i,j\rangle_{\text{dist}} = |j,i\rangle_{\text{dist}}$. The normalized bosonic basis states are
\begin{equation}\label{eq:symmetrization}
|i,j\rangle_{\text{bos}} =
\begin{cases}
|i,i\rangle_{\text{dist}}, & i=j,\\
\frac{1}{\sqrt{2}}(|i,j\rangle_{\text{dist}} + |j,i\rangle_{\text{dist}}), & i<j.
\end{cases}
\end{equation}
Since $P_{\text{sym}}$ commutes with $U_g^{\text{dist}}$ (as $U_g^{\text{dist}}$ preserves particle exchange symmetry), the symmetry $U_g^{\text{dist}}$ descends to a well-defined symmetry $U_g^{\text{bos}}$ on $\mathcal{H}_2^{\text{bos}}$. Consequently, any symmetry of the original single-particle lattice induces a corresponding symmetry in the many-body Fock-space lattice.

The many-body evolution paths are the geometric paths on the Fock-space lattice, and their symmetry relations (e.g., pairwise cancellation or conjugate-phase relations) usually follow directly from those of the distinguishable-particle tensor-product lattice after symmetrization. In this sense, the many-body EPS is not merely an artificial construction but a rigorous consequence of extending single-particle lattice symmetries to the Fock space via symmetrization.

For example, in Fig.~\ref{fig:paths}, the EPS between the upper and lower path groups arises from reflection symmetry between the B and C sites combined with complex conjugation. The complex conjugate has no effect here, since the accumulation phases of the evolution paths are either $0$ or $\pi$. While the cancellation within each group of evolution paths (e.g., path 1 to 5: -2,2,-4,2,2) looks a bit strange but it is also due to the symmetry between the two evolution path surrounding the plaque with $\pi$ flux. We illustrate the product of the rhombic lattice for two distinguishable particles in Fig.~\ref{fig:paths_cancellation}. The two paths denoted by red and blue dashed lines have opposite contributions, because they can be viewed as the product of the same first half section on the ``small'' rhombic lattice and opposite latter half sections on the ``big'' rhombic lattice. The symmetrization of these pairwise paths naturally leads to the cancellation among the evolution paths for indistinguishable particles shown in Fig.~\ref{fig:paths}, as symmetrization can be seen as the recombination of intermediate states (see Eq.~(\ref{eq:symmetrization})) and thus neither adds nor removes the contribution of any path (see Appendix A for a detailed proof).

Interactions introduce occupation-dependent on-site phase shifts for evolution paths, which may breaks EPS. The on-site interaction term $H_{\text{int}} = \frac{U}{2} \sum_i n_i (n_i - 1)$ contributes a phase factor $e^{-i U n_i (n_i-1) \Delta t / 2}$ for a site with occupation $n_i$ over a time step $\Delta t$. Since different paths may visit sites with different occupation histories, the combination of paths that were symmetric in the non-interacting limit may now accumulate different interaction-induced phases. This breaks the precise phase relationship, lifting the destructive interference cancellation and thereby enabling transport.
}

\section{EPS breaking and interaction-induced chiral transports in a flux ladder}\label{sec:ladder}

\begin{figure*}[t]
\centering
\includegraphics[width=1\linewidth]{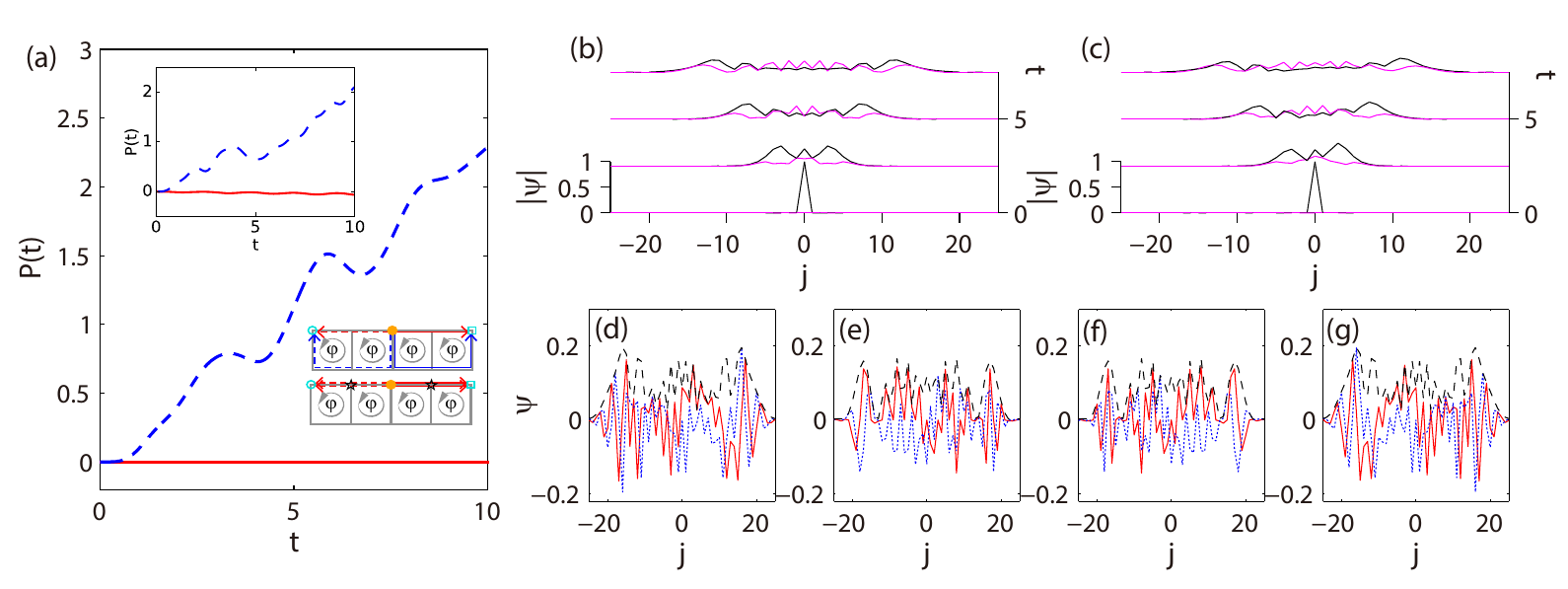}
\caption{(a) Time evolution of the chiral displacement $P(t)$ for a BEC initial state at site $\hat{a}_{0}$ on the flux ladder. For $U=0$ (red solid), $P(t)$ remains zero due to conjugate-phase EPS. For $U=4$ (blue dashed), the EPS is broken, leading to a finite chiral current. {Higher left inset: The chiral displacement of a two-particle initial state $|0a\rangle \bigotimes |0a\rangle$ calculated with density matrix renormalization group (DMRG) method.} Lower right inset: Schematic of the flux ladder and examples of symmetric single-particle paths (solid and dashed lines) that are conjugates when $U=0$. It shows how an interaction event (pentagram) breaks the conjugate-phase relationship for a pair of two-particle paths. (b)(c): The time evolution of the upper (black) and lower (pink) components in the free and interacting cases are shown in (b) and (c), respectively.  (d)(e): the wave functions of the upper (d) and lower (e) components at $t=10$ for the free case. (f)(g): the wave functions of the upper (f) and lower (g) components at $t=10$ for the interacting case. The red solid, blue dotted and black dashed curves represent the real part, imaginary part and module of the wave functions, respectively. The spatial distributions of the wave function amplitudes are symmetric in the noninteracting case (see panels (b), (d) and (e)), but become slightly asymmetric in the interacting case (panels (c), (f) and (g)). {Here we set $J=1$ as the unit of energy. The system size is $L=51$, and open boundary conditions are employed. The flux phase is $\varphi=\pi/3$. We use the mean-field approximation and the Gross-Pitaevskii equation to calculate the dynamical evolution.} }
\label{fig:HHM}
\end{figure*}

We next apply the EPS framework to explain interaction-induced chiral transports in a flux ladder, demonstrating its generality beyond the AB cage system. Consider a two-leg Harper-Hofstadter ladder described by
\begin{equation}
\begin{split}
H = & J \sum_j \left( e^{i\frac{\varphi}{2}} \hat{a}_j^\dagger \hat{a}_{j+1} + e^{-i\frac{\varphi}{2}} \hat{b}_j^\dagger \hat{b}_{j+1} + \hat{b}_j^\dagger \hat{a}_j + \text{H.c.} \right) \\
& + \frac{1}{2} U \sum_{j,\xi=a,b} \hat{n}_{j\xi} (\hat{n}_{j\xi} - 1),
\end{split}
\label{eq:HH_ladder}
\end{equation}
where $\hat{a}_j, \hat{b}_j$ are annihilation operators on the two legs. This model possesses a chiral symmetry: $U_{\text{ch}} \hat{a}_j U_{\text{ch}}^\dagger = \hat{b}_{-j}$ and $U_{\text{ch}} \hat{b}_j U_{\text{ch}}^\dagger = \hat{a}_{-j}$.

We assume a particle is prepared on the central site of leg $a$, i.e., the initial state is $|0a\rangle$. On one hand, the evolution paths from $|0a\rangle$ to $|ja\rangle$ and $|-ja\rangle$ are conjugate symmetric, i.e., $\mathcal{P}_{n}^{(\xi)}(j\lambda,0a)=\mathcal{P}_{n}^{(\xi)*}(-j\lambda,0a)$ with $\lambda=a, b$, which is shown with the solid and dashed lines in the upper subfigure of the inset of Fig.~\ref{fig:HHM}(a). On the other hand, in the expansion of the propagator (Eq.~(\ref{eq:path_expansion})), either the odd-order or the even-order terms are nonvanishing, because any evolution path connecting the initial site and the final site is longer than the shortest path with an even number of hopping steps. If the shortest paths involve an even (odd) number of hopping steps, paths with an even (odd) number of hopping steps are possible. Note that the coefficient $\left(-\frac{i\Delta t}{\hbar}\right)^{N}$ is purely real when $N$ is even and purely imaginary when $N$ is odd. Therefore, the propagators satisfy $|K(j\lambda,t;0a,t_0)| = |K(-j\lambda,t;0a,t_0)|$, and the chiral displacement $\hat{P} = \sum_{j} j (\hat{a}^\dagger_j \hat{a}_j - \hat{b}^\dagger_j \hat{b}_j)$ has a vanishing expectation value, because the probability distribution is always symmetric with respect to the central site.

In contrast, the evolution paths of a many-body state with contact interactions even becomes unsymmetrical. For example, for the simplest case with two-particle state, if $U=0$, the phase factors of symmetric two-particle paths are still conjugate to each other. But, when $U\neq 0$,  on the crossing which indicate more than one particles occupy the same site (see the pentagrams in the lower subfigure of the inset of Fig.~\ref{fig:HHM}(a)),  the contribution of interaction for two symmetric two-particle paths are not conjugate to each other. In this case, the path symmetry is broken. Therefore, we no longer have the condition $\mathcal{P}_{n}^{(\xi)}(j\lambda,0a)=\mathcal{P}_{n}^{(\xi)*}(-j\lambda,0a)$ and $P_{|a0\rangle}(t)=P_{|b0\rangle}(t)=0$, when the interaction term is turned on. This means that the propagation of a local state becomes asymmetric. Together with the intrinsic chiral symmetry of the model, which implies that the center-of-mass propagation directions of the two legs are opposite, this asymmetric propagation leads to chiral transport.

The above discussion is also true for a BEC initial state that is more easily achieved in experiment, since coherent state can be seen as the summation of Fock states. Our numerical calculation with Gross-Pitaevskii (GP) equation completely supports the above analysis. As shown in Fig.~\ref{fig:HHM}(a), the based density in the free case is always zero, but it in the interacting case is finite. In fact, interaction-induced chiral currents of a two-particle local initial state has been observed in experiment~\cite{tai2017microscopy}. The only difference is that, the initial state there is a two-particle product state of $|a0\rangle$ and $|b0\rangle$. In this case, interactions also break the symmetry between the two evolution paths connecting the initial states and two objective spatially symmetric points with respect to $j=0$ on both legs, thereby giving rise to chiral transport. In contrast to the initial state considered above, here the initial state is symmetric between the two legs, the time-evolved wave functions even satisfy chiral symmetry.

\section{Conclusion and Discussion}\label{sec:conclusion}

We have developed the concept of evolution-path symmetry (EPS) as an alternative framework for understanding interaction-induced transport phenomena in flux lattices. Starting from a concrete example of the $\pi$-flux rhombic lattice, we showed how interactions break the destructive interference between symmetric paths, enabling delocalization beyond the AB cage. This observation led us to formalize the notion of EPS and classify its two key types: destructive-interference EPS, which governs localization, and conjugate-phase EPS, which suppresses chiral currents in non-interacting systems.

Applying this framework to flux ladders, we demonstrated how interactions break the unchanged-/conjugate-phase EPS, leading to finite chiral currents. In both systems, the breaking occurs because interactions introduce occupation-dependent phase shifts that disrupt the precise phase relationships between symmetry-related paths.

The EPS framework shifts the focus from static microscopic properties (i.e., eigenstate spectra, energy bands and initial-state projection) to the dynamical interference of trajectories in many-body space. It clarifies why interactions, which introduce non-linear phase accumulation dependent on local occupation, are generically effective at breaking symmetries that rely on precise phase relationships between paths. {This unified framework also works for particles wiith other different quantum statistics, such as fermions and hard-core bosons (see Appendix B for more details).}

{
Our theoretical framework is directly relevant to several recent experimental advances. The interaction-induced chiral transport in flux ladders has been observed in ultracold atomic gases using quantum gas microscopy~\cite{tai2017microscopy}, where the two-body dynamics in a Harper-Hofstadter ladder were directly imaged. More recently, interaction-induced asymmetric chiral currents have been reported in a synthetic zigzag ladder of ultracold atoms~\cite{Zhao2025}. In the context of flat-band localization, the Aharonov-Bohm caging and its interaction-induced breakdown have been experimentally demonstrated in superconducting qubit arrays~\cite{martinez2023flat} and in Rydberg-atom synthetic lattices~\cite{chen2025interaction}. Our EPS framework provides a unified dynamical explanation for these observations, showing that they all originate from the breaking of path symmetries due to interaction-induced phase shifts. This perspective not only connects seemingly distinct experimental phenomena but also offers a predictive tool for designing future experiments in other lattice geometries and interaction regimes.
}

We emphasize that this paper discusses only localized initial states. The analysis becomes considerably more complex if nonlocal initial states are considered. Although the analysis of evolution paths is independent of the initial state, the actual dynamical evolution depends critically on it. Furthermore, cases involving time-dependent Hamiltonians or non-conservation of particle number also present significant analytical challenges. Consequently, our theory provides only a preliminary framework for understanding quantum transport phenomena through the lens of evolution-path symmetry.

\begin{acknowledgments}
This work is supported by the National Natural Science Foundation of China (NSFC) under Grant Nos.~12574297, 12174233, 12004230, 12034012 and 12374479, the Natural Science Foundation of Sichuan Province under Grant No. 2025ZNSFSC0058, the Science Specialty Program of Sichuan University under Grant No. 2020SCUNL210, the Fundamental Research Funds for the Central Universities under Grant No. YJ202212, the Fundamental Research Program of Shanxi Province under Grant No. 202403021221024, and the National Key R$\&$D Program of China under Grants No. 2024YFF0508503. WY acknowledges support from the and Quantum Science and Technology-National Science and Technology Major Project (Grant No. 2021ZD0301904).
\end{acknowledgments}

{
\section*{Appendix A: Symmetrization and many-body path amplitudes}

In this supplemental material, we provide a detailed derivation of how the symmetrization of distinguishable-particle Fock states leads to the path amplitudes in the indistinguishable-particle Fock-space lattice. The key point is that symmetrization does not add or remove any path; it merely recombines the contributions from distinguishable-particle paths to form the effective amplitudes for indistinguishable particles.

\subsection*{Notation and setup}

We consider two identical bosons on a lattice. The Hilbert space for distinguishable particles is $\mathcal{H}_2^{\text{dist}} = \mathcal{H}_1 \otimes \mathcal{H}_1$, with orthonormal basis $|i,j\rangle_{\text{dist}} = |i\rangle \otimes |j\rangle$. The Hamiltonian for distinguishable particles is $H_{\text{dist}} = h \otimes I + I \otimes h$, where $h$ is the single-particle Hamiltonian with hopping amplitude $t$ (set to $1$ in the following for simplicity).

The symmetric (bosonic) subspace is obtained via the projection operator $P_{\text{sym}} = \frac{1}{2}(I + \text{SWAP})$, where $\text{SWAP}|i,j\rangle_{\text{dist}} = |j,i\rangle_{\text{dist}}$. The normalized bosonic basis states are
\begin{equation}
|i,j\rangle_{\text{bos}} =
\begin{cases}
|i,i\rangle_{\text{dist}}, & i = j,\\[4pt]
\dfrac{1}{\sqrt{2}}\bigl(|i,j\rangle_{\text{dist}} + |j,i\rangle_{\text{dist}}\bigr), & i < j.
\end{cases}
\end{equation}

The bosonic Hamiltonian is $H_{\text{bos}} = P_{\text{sym}} H_{\text{dist}} P_{\text{sym}}$ restricted to the symmetric subspace. For simplicity, we only consider a Hamiltonian with a constant hopping coefficient between nearest-neighbor sites.

\subsection*{Case 1: Non-doubly occupied initial state, doubly occupied final state}

Let $i < j$. The initial state is $|i,j\rangle_{\text{bos}}$ and the final state is $|j,j\rangle_{\text{bos}}$. Expanding in the distinguishable basis:
\begin{equation}
|i,j\rangle_{\text{bos}} = \frac{1}{\sqrt{2}}\bigl(|i,j\rangle_{\text{dist}} + |j,i\rangle_{\text{dist}}\bigr),\qquad
|j,j\rangle_{\text{bos}} = |j,j\rangle_{\text{dist}}.
\end{equation}

The matrix element is
\begin{equation}
\langle j,j | H_{\text{bos}} | i,j \rangle_{\text{bos}} = \langle j,j |_{\text{dist}} H_{\text{dist}} \left[ \frac{1}{\sqrt{2}}\bigl(|i,j\rangle_{\text{dist}} + |j,i\rangle_{\text{dist}}\bigr) \right].
\end{equation}

Consider the hopping term $-t|j\rangle\langle i| \otimes I$ (first particle hops from $i$ to $j$): (1) Acting on $|i,j\rangle_{\text{dist}}$: $-t|j,j\rangle_{\text{dist}}$. (2) Acting on $|j,i\rangle_{\text{dist}}$: the first particle is at $j$, not $i$, so the contribution is $0$.
Thus from this term we get $\frac{1}{\sqrt{2}}(-t|j,j\rangle_{\text{dist}})$, and the inner product with $\langle j,j|_{\text{dist}}$ gives $-t/\sqrt{2}$.

Now consider the term $-t I \otimes |j\rangle\langle i|$ (second particle hops from $i$ to $j$): (1) Acting on $|i,j\rangle_{\text{dist}}$: the second particle is at $j$, not $i$, so contribution is $0$. (2) Acting on $|j,i\rangle_{\text{dist}}$: the second particle is at $i$, hops to $j$, giving $-t|j,j\rangle_{\text{dist}}$. This contributes $\frac{1}{\sqrt{2}}(-t|j,j\rangle_{\text{dist}})$, and the inner product again gives $-t/\sqrt{2}$.

Adding both contributions:
\begin{equation}
\langle j,j | H_{\text{bos}} | i,j \rangle_{\text{bos}} = -\frac{t}{\sqrt{2}} - \frac{t}{\sqrt{2}} = -\sqrt{2}\,t \qquad (i < j).
\end{equation}

\subsection*{Case 2: Doubly occupied initial state, non-doubly occupied final state}

Let $i < j$. The initial state is $|i,i\rangle_{\text{bos}} = |i,i\rangle_{\text{dist}}$, and the final state is
\begin{equation}
|i,j\rangle_{\text{bos}} = \frac{1}{\sqrt{2}}\bigl(|i,j\rangle_{\text{dist}} + |j,i\rangle_{\text{dist}}\bigr).
\end{equation}

Take the hopping term $-t|j\rangle\langle i| \otimes I$ (first particle hops from $i$ to $j$): Acting on $|i,i\rangle_{\text{dist}}$ leads to $-t|j,i\rangle_{\text{dist}}$. Inner product with the final state:
\begin{equation}
\langle i,j|_{\text{dist}}(-t|j,i\rangle_{\text{dist}}) = 0,\qquad
\langle j,i|_{\text{dist}}(-t|j,i\rangle_{\text{dist}}) = -t.
\end{equation}
Multiplying by the final-state coefficient $1/\sqrt{2}$ gives $-t/\sqrt{2}$.

Now consider the term $-t I \otimes |j\rangle\langle i|$ (second particle hops from $i$ to $j$): Acting on $|i,i\rangle_{\text{dist}}$ leads to $-t|i,j\rangle_{\text{dist}}$. Inner product:
\begin{equation}
\langle i,j|_{\text{dist}}(-t|i,j\rangle_{\text{dist}}) = -t,\qquad
\langle j,i|_{\text{dist}}(-t|i,j\rangle_{\text{dist}}) = 0.
\end{equation}
Again multiplying by $1/\sqrt{2}$ gives $-t/\sqrt{2}$.

Adding both gives rise to
\begin{equation}
\langle i,j | H_{\text{bos}} | i,i \rangle_{\text{bos}} = -\frac{t}{\sqrt{2}} - \frac{t}{\sqrt{2}} = -\sqrt{2}\,t \qquad (i < j).
\end{equation}

\subsection*{Case 3: Non-doubly occupied initial and final states, single path}

Consider initial state $|i,j\rangle_{\text{bos}}$ ($i<j$) and final state $|k,j\rangle_{\text{bos}}$ ($k<j$), with $k$ a nearest neighbor of $i$ but not of $j$. Only one path contributes (the first particle hops from $i$ to $k$), and a direct calculation (similar to the above but without the second-path contribution) yields
\begin{equation}
\langle k,j | H_{\text{bos}} | i,j \rangle_{\text{bos}} = -t.
\end{equation}

\subsection*{Case 4: Two-path interference}

Now consider initial state $|i,j\rangle_{\text{bos}}$ ($i<j$) and final state $|k,j\rangle_{\text{bos}}$ ($k<j$), with $k$ being a nearest neighbor of both $i$ and $j$. Expanding:
\begin{equation}
|k,j\rangle_{\text{bos}} = \frac{1}{\sqrt{2}}\bigl(|k,j\rangle_{\text{dist}} + |j,k\rangle_{\text{dist}}\bigr).
\end{equation}
(1) From $|i,j\rangle_{\text{dist}}$, first particle $i \to k$ gives $|k,j\rangle_{\text{dist}}$. Coefficient product: $(1/\sqrt{2}) \cdot (-t) \cdot (1/\sqrt{2}) = -t/2$.
(2) From $|j,i\rangle_{\text{dist}}$, second particle $i \to k$ gives $|j,k\rangle_{\text{dist}}$. Coefficient product: $(1/\sqrt{2}) \cdot (-t) \cdot (1/\sqrt{2}) = -t/2$.
The total contribution is
\begin{equation}
-\frac{t}{2} + \left(-\frac{t}{2}\right) = -t,
\end{equation}
rather than $-2t$. The factor $1/2$ from the product of normalization coefficients prevents the emergence of a $-2t$ coefficient.

\subsection*{Summary of matrix elements}

\begin{equation}
\begin{array}{lll}
\hline
\text{Initial state} & \text{Final state} & \text{Matrix element} \\
\hline
|i,j\rangle_{\text{bos}}\; & |k,j\rangle_{\text{bos}}\;(k<j,\; \text{single path}) & -t \\
|i,j\rangle_{\text{bos}}\; & |j,j\rangle_{\text{bos}} & -\sqrt{2}\,t \\
|i,i\rangle_{\text{bos}} & |i,k\rangle_{\text{bos}}\;(i<k) & -\sqrt{2}\,t \\
|i,j\rangle_{\text{bos}}\; & |k,j\rangle_{\text{bos}}\;( \text{two paths}) & -t \\
\hline
\end{array}
\end{equation}

Thus, the many-body Fock-space lattice contains only two types of hopping coefficients: $-t$ and $-\sqrt{2}\,t$. A crucial insight is that the $\sqrt{2}$ factor is not an anomaly that disrupts path contribution; rather is precisely the factor that restores the correct quantum statistics when transitioning from a distinguishable-particle picture to an indistinguishable one.

Consider a process connecting two doubly occupied states, say $|i,i\rangle_{\text{bos}} \to |j,j\rangle_{\text{bos}}$ via a path that goes through an intermediate non-doubly occupied state $|i,j\rangle_{\text{bos}}$. In the normalized bosonic Fock basis, each hopping step involving a change from a doubly occupied site to a non-doubly occupied site (or vice versa) carries a factor $\sqrt{2}$. For instance, the amplitude for $|i,i\rangle_{\text{bos}} \to |i,j\rangle_{\text{bos}} \to |j,j\rangle_{\text{bos}}$ is $(\sqrt{2}\,t) \cdot (\sqrt{2}\,t) = 2t^{2}$ (up to phases).

Now, if we forget particle indistinguishability and treat the particles as distinguishable, the same physical process would be represented by two distinct paths in the distinguishable-particle Hilbert space:
\begin{enumerate}
\item Path A: Particle 1 goes $i \to j$, particle 2 stays at $i$, then particle 2 goes $i \to j$.
\item Path B: Particle 2 goes $i \to j$, particle 1 stays at $i$, then particle 1 goes $i \to j$.
\end{enumerate}
In the distinguishable-particle picture, each of these paths contributes an amplitude $t^{2}$ (ignoring phases). Their total contribution is $t^{2} + t^{2} = 2t^{2}$, which exactly matches the $2t^{2}$ obtained from the bosonic calculation. The $\sqrt{2}$ factor per hopping step therefore compensates for the doubling of paths that occurs when we ``unlabel'' the particles.

Generalizing this logic: for any path connecting two doubly occupied states in the bosonic Fock-space lattice, each time the path enters or exits a doubly occupied site, a factor $\sqrt{2}$ appears. When we ``unravel'' this bosonic path into the corresponding distinguishable-particle paths, each such $\sqrt{2}$ factor accounts for the combinatorial multiplicity of ways that identical particles can be assigned to the two available labels. Importantly, this compensation is exact and local---it applies to each segment of the path independently.

Consequently, the cancellation among different bosonic paths (e.g., the vanishing sum $-2+2-4+2+2 = 0$ in Fig.~\ref{fig:paths}) is directly inherited from the distinguishable-particle calculation. The $\sqrt{2}$ factors do not disrupt the cancellation; they are precisely the factors that make the bosonic path amplitudes equal to the sum of their distinguishable-particle counterparts. Thus, the appearance of $\sqrt{2}$ is a feature, not a bug, of the bosonic representation.

\subsection*{Relation to the main text}

The above derivation shows that symmetrization of the distinguishable-particle Fock-state lattice does not introduce or remove any path; it merely reassembles the contributions from distinguishable-particle paths into effective amplitudes for indistinguishable particles. The $\sqrt{2}$ factor that appears in hopping coefficients involving doubly occupied states is not merely a normalization convention but a direct consequence of bosonic statistics: it accounts for the combinatorial multiplicity of label assignments when particles are treated as indistinguishable. The cancellation observed in Fig.~\ref{fig:paths} of the main text (e.g., the vanishing sum $-2+2-4+2+2 = 0$) is a direct consequence of this symmetrization procedure combined with the $\pi$-flux condition in the original rhombic lattice. The EPS framework thus provides a systematic way to understand many-body interference phenomena by tracing them back to the symmetries of the underlying distinguishable-particle tensor-product lattice, where each bosonic path amplitude is naturally interpreted as the coherent sum over all labelings of the distinguishable-particle paths.

}

{
\section*{Appendix B: Extension to fermions and hard-core bosons}
In this appendix, we discuss how the EPS framework extends to other quantum statistics, namely fermions and hard-core bosons. The key point is that the path symmetry argument does not rely on the details of quantum statistics in an essential way. The same single-particle symmetry structure also exists in the fermionic case, as well as at the distinguishable-particle level. The only difference is that when we go from distinguishable particles to indistinguishable fermions, we need to use anti-symmetrization instead of symmetrization.

\subsection*{Anti-symmetrization for fermions}

For two identical fermions, the anti-symmetrized two-fermion state is written as
\begin{equation}
|i,j\rangle_{\text{fer}} = \frac{1}{\sqrt{2}}\bigl(|i,j\rangle_{\text{dist}} - |j,i\rangle_{\text{dist}}\bigr), \qquad i<j.
\end{equation}

We now examine how the hopping processes in the distinguishable-particle Fock-space lattice correspond to those in the indistinguishable-fermion Fock-space lattice. There are only a few cases (see Appendix A for details).

\paragraph*{Case 1: Single hopping channel.}
Two particles occupy different sites for both the initial and final states, and there is only one hopping channel: $|i,j\rangle_{\text{dist}} \to |k,j\rangle_{\text{dist}}$, with $j>i$, $k>i$, and only $i$ can hop to $k$. In this case, the transition $|i,j\rangle_{\text{fer}} \to |k,j\rangle_{\text{fer}}$ corresponds to two paths. One is the first particle hopping: $|i,j\rangle_{\text{dist}} \to |k,j\rangle_{\text{dist}}$, contributing $+1/2$. The other is the second particle hopping: $-|j,i\rangle_{\text{dist}} \to -|j,k\rangle_{\text{dist}}$, also contributing $+1/2$, not $-1/2$.

\paragraph*{Case 2: Two hopping channels.}
Two particles occupy different sites for both the initial and final states, and there are two hopping channels: $|i,j\rangle_{\text{dist}} \to |k,j\rangle_{\text{dist}}$ and $|i,j\rangle_{\text{dist}} \to |i,k\rangle_{\text{dist}}$, with $j>i$, $k>i$, and both $i$ and $j$ can hop to $k$. In this case, the transition $|i,j\rangle_{\text{fer}} \to |i,k\rangle_{\text{fer}}$ again corresponds to two paths: first particle hopping ($|i,j\rangle_{\text{dist}} \to |k,j\rangle_{\text{dist}}$) contributing $+1/2$, and second particle hopping ($-|j,i\rangle_{\text{dist}} \to -|j,k\rangle_{\text{dist}}$) also contributing $+1/2$. The path where the particle at $j$ hops to $k$ does not contribute here because the final state would be $|i,k\rangle_{\text{fer}}$, not $|k,j\rangle_{\text{fer}}$.

\paragraph*{Case 3: Double occupancy.}
When the initial or final state involves double occupancy, e.g., $|i,i\rangle_{\text{dist}}$, such a state exists for distinguishable particles but does not exist for fermions after anti-symmetrization. The corresponding anti-symmetrized two-fermion state would be
\begin{equation}
\frac{1}{\sqrt{2}}\bigl(|i,i\rangle_{\text{dist}} - |i,i\rangle_{\text{dist}}\bigr) = 0.
\end{equation}
Nevertheless, we can still write the transition processes as combinations of distinguishable-particle paths with opposite signs, such as $|i,j\rangle_{\text{dist}} \to |i,i\rangle_{\text{dist}}$, $-|j,i\rangle_{\text{dist}} \to |i,i\rangle_{\text{dist}}$, $|i,j\rangle_{\text{dist}} \to -|i,i\rangle_{\text{dist}}$, and $-|j,i\rangle_{\text{dist}} \to -|i,i\rangle_{\text{dist}}$, all of which cancel out. Thus the total transition amplitude is zero.

\subsection*{Implications for EPS}
In summary, except for case 3, the hopping processes on the indistinguishable-particle Fock lattice can be mapped back to a pair of hopping processes on the distinguishable-particle Fock lattice with the same factors. The pair of hopping processes are connected by the exchange of particles. This conclusion is the same for bosons. On the other hand, paths that involve double occupancy have no contribution. Therefore, the EPS for fermions is similar to that for bosons discussed above, except that evolution involving double occupation vanishes (meaning it does not violate any symmetry). We can conclude that quantum statistics does not essentially affect EPS, and the phenomena predicted by EPS (e.g., the interaction-induced transport discussed in this paper) still hold for fermions.

For example, for the rhombic lattice with $\pi$ flux, we can find that pairwise paths cancel each other by mapping back to the distinguishable-particle Fock-state lattice, which is simply the product of single-particle rhombic lattices. Therefore, the cancellation among evolution paths still holds for distinguishable particles.

\subsection*{Hard-core bosons}
For hard-core bosons, the symmetrization procedure is similar to that for soft-core bosons, except that double occupancy is effectively forbidden (or treated as having infinite repulsion). A similar analysis shows that quantum statistics does not affect EPS, and the phenomena predicted by EPS also hold for hard-core bosons.

\subsection*{Finite-range interactions}
As for finite-range interactions, we also believe they are not essential for the main predictions of this paper, but the details need to be analyzed on a case-by-case basis. For example, for fermions, a next-nearest-neighbor interaction would definitely break the EPS for the rhombic lattice with $\pi$ flux due to the interaction-induced the same (not opposite) phase accumulations.

\subsection*{Conclusion}
Quantum statistics does not fundamentally alter the path symmetry picture, but each case requires a careful treatment of the (anti-)symmetrization factors. The EPS framework thus provides a unified approach that can be extended beyond soft-core bosons to fermions and hard-core bosons, as well as to systems with finite-range interactions, provided that the symmetry analysis is performed carefully.
}


\begin{thebibliography}{99}

\bibitem{anderson1958absence}
P. W. Anderson,
\newblock Absence of diffusion in certain random lattices,
\newblock \textit{Phys. Rev.} \textbf{109}, 1492 (1958).

\bibitem{brandes2003the}
T. Brandes and S. Kettemann,
\newblock \textit{The Anderson Transition and its Ramifications --- Localisation, Quantum Interference, and Interactions},
\newblock Lecture Notes in Physics, Vol. 630 (Springer, Berlin, 2003), ISBN 978-3-642-07398-4.

\bibitem{Wiersma1997}
D. S. Wiersma, P. Bartolini, A. Lagendijk, and R. Righini,
\newblock Localization of light in a disordered medium,
\newblock \textit{Nature} \textbf{390}, 671 (1997).

\bibitem{Storzer2006}
M. St\"orzer, P. Gross, C. M. Aegerter, and G. Maret,
\newblock Observation of the critical regime near Anderson localization of light,
\newblock \textit{Phys. Rev. Lett.} \textbf{96}, 063904 (2006).

\bibitem{Scheffold1999}
F. Scheffold, R. Lenke, R. Tweer, and G. Maret,
\newblock Localization or classical diffusion of light?,
\newblock \textit{Nature} \textbf{398}, 206 (1999).

\bibitem{Schwartz2007}
T. Schwartz, G. Bartal, S. Fishman, and M. Segev,
\newblock Transport and Anderson localization in disordered two-dimensional photonic lattices,
\newblock \textit{Nature} \textbf{446}, 52 (2007).

\bibitem{Lahini2008}
Y. Lahini, A. Avidan, F. Pozzi, M. Sorel, R. Morandotti, D. N. Christodoulides, and Y. Silberberg,
\newblock Anderson localization and nonlinearity in one-dimensional disordered photonic lattices,
\newblock \textit{Phys. Rev. Lett.} \textbf{100}, 013906 (2008).

\bibitem{Karbasi2012}
S. Karbasi, C. R. Mirr, P. G. Frazier, R. J. Frazier, and A. Mafi,
\newblock Observation of transverse Anderson localization in an optical fiber,
\newblock \textit{Opt. Lett.} \textbf{37}, 2304 (2012).

\bibitem{Karbasi2014}
S. Karbasi, K. W. Koch, and A. Mafi,
\newblock Image transport through a disordered optical fibre mediated by transverse Anderson localization,
\newblock \textit{Nat. Commun.} \textbf{5}, 3362 (2014).

\bibitem{Billy2008}
J. Billy, V. Josse, Z. Zuo, A. Bernard, B. Hambrecht, P. Lugan, D. Cl\'ement, L. Sanchez-Palencia, P. Bouyer, and A. Aspect,
\newblock Direct observation of Anderson localization of matter waves in a controlled disorder,
\newblock \textit{Nature} \textbf{453}, 891 (2008).

\bibitem{Roati2008}
G. Roati, C. D'Errico, L. Fallani, M. Fattori, C. Fort, M. Zaccanti, G. Modugno, M. Modugno, and M. Inguscio,
\newblock Anderson localization of a non-interacting Bose-Einstein condensate,
\newblock \textit{Nature} \textbf{453}, 895 (2008).

\bibitem{Ludlam2005}
J. J. Ludlam, S. N. Taraskin, and S. R. Elliott,
\newblock Universal features of localized eigenstates in disordered systems,
\newblock \textit{J. Phys.: Condens. Matter} \textbf{17}, L321 (2005).

\bibitem{Conti2008}
C. Conti and A. Fratalocchi,
\newblock Dynamic light diffusion, three-dimensional Anderson localization and lasing in inverted opals,
\newblock \textit{Nat. Phys.} \textbf{4}, 794 (2008).

\bibitem{Hu2008}
H. Hu, A. Strybulevych, J. H. Page, S. E. Skipetrov, and B. A. van Tiggelen,
\newblock Localization of ultrasound in a three-dimensional elastic network,
\newblock \textit{Nat. Phys.} \textbf{4}, 945 (2008).

\bibitem{Chabe2008}
J. Chab\'e, G. Lemari\'e, B. Gr\'emaud, D. Delande, P. Szriftgiser, and J. C. Garreau,
\newblock Experimental observation of the Anderson metal-insulator transition with atomic matter waves,
\newblock \textit{Phys. Rev. Lett.} \textbf{101}, 255702 (2008).

\bibitem{Ying2016}
T. Ying, Y. Chen, Y. Wang, and X. Chen,
\newblock Anderson localization of electrons in single crystals: Li$_x$Fe$_7$Se$_8$,
\newblock \textit{Sci. Adv.} \textbf{2}, e1501283 (2016).

\bibitem{Choi2018}
S. H. Choi, S. W. Kim, Z. Ku, M. A. Visbal-Onufrak, S. R. Kim, K. H. Choi, H. Ko, W. Choi, A. M. Urbas, T. W. Goo, Y. L. Kim, and S. H. Kim,
\newblock Anderson light localization in biological nanostructures of native silk,
\newblock \textit{Nat. Commun.} \textbf{9}, 452 (2018).

\bibitem{Skipetrov2016}
S. E. Skipetrov and I. M. Sokolov,
\newblock Red light for Anderson localization,
\newblock \textit{New J. Phys.} \textbf{18}, 021001 (2016).

\bibitem{Aharonov1959}
Y. Aharonov and D. Bohm,
\newblock Significance of electromagnetic potentials in the quantum theory,
\newblock \textit{Phys. Rev.} \textbf{115}, 485 (1959).

\bibitem{vidal1998aharonov}
J. Vidal, R. Mosseri, and B. Dou\c{c}ot,
\newblock Aharonov-Bohm cages in two-dimensional structures,
\newblock \textit{Phys. Rev. Lett.} \textbf{81}, 5888 (1998).

\bibitem{Abilio1999}
C. C. Abilio, P. Butaud, Th. Fournier, B. Pannetier, J. Vidal, S. Tedesco, and B. Dalzotto,
\newblock Magnetic field induced localization in a two-dimensional superconducting wire network,
\newblock \textit{Phys. Rev. Lett.} \textbf{83}, 5102 (1999).

\bibitem{Naud2001}
C. Naud, G. Faini, and D. Mailly,
\newblock Aharonov-Bohm cages in 2D normal metal networks,
\newblock \textit{Phys. Rev. Lett.} \textbf{86}, 5104 (2001).

\bibitem{Li2022}
H. Li, Z. Dong, S. Longhi, Q. Liang, D. Xie, and B. Yan,
\newblock Aharonov-Bohm caging and inverse Anderson transition in ultracold atoms,
\newblock \textit{Phys. Rev. Lett.} \textbf{129}, 220403 (2022).

\bibitem{Li2025}
H. Li, Q. Liang, Z. Dong, H. Wang, W. Yi, J.-S. Pan, and B. Yan,
\newblock Engineering topological chiral transport in a flat-band lattice of ultracold atoms,
\newblock \textit{Light Sci. Appl.} \textbf{14}, 326 (2025).

\bibitem{Nandkishore2015}
R. Nandkishore and D. A. Huse,
\newblock Many-body localization and thermalization in quantum statistical mechanics,
\newblock \textit{Annu. Rev. Condens. Matter Phys.} \textbf{6}, 15 (2015).


\bibitem{Fleishman1980}
L. Fleishman and P. W. Anderson,
\newblock Interactions and the Anderson transition,
\newblock \textit{Phys. Rev. B} \textbf{21}, 2366 (1980).

\bibitem{Altshuler1997}
B. L. Altshuler, Y. Gefen, A. Kamenev, and L. S. Levitov,
\newblock Quasiparticle lifetime in a finite system: a nonperturbative approach,
\newblock \textit{Phys. Rev. Lett.} \textbf{78}, 2803 (1997).

\bibitem{Gornyi2005}
I. V. Gornyi, A. D. Mirlin, and D. G. Polyakov,
\newblock Interacting electrons in disordered wires: Anderson localization and low-T transport,
\newblock \textit{Phys. Rev. Lett.} \textbf{95}, 206603 (2005).

\bibitem{Basko2006}
D. M. Basko, I. L. Aleiner, and B. L. Altshuler,
\newblock Metal-insulator transition in a weakly interacting many-electron system with localized single-particle states,
\newblock \textit{Ann. Phys.} \textbf{321}, 1126 (2006).

\bibitem{Oganesyan2007}
V. Oganesyan and D. A. Huse,
\newblock Localization of interacting fermions at high temperature,
\newblock \textit{Phys. Rev. B} \textbf{75}, 155111 (2007).

\bibitem{Pal2010}
A. Pal and D. A. Huse,
\newblock Many-body localization phase transition,
\newblock \textit{Phys. Rev. B} \textbf{82}, 174411 (2010).

\bibitem{Imbrie2016}
J. Z. Imbrie,
\newblock On many-body localization for quantum spin chains,
\newblock \textit{J. Stat. Phys.} \textbf{163}, 998 (2016).

\bibitem{vidal2000interaction}
J. Vidal, B. Dou\c{c}ot, R. Mosseri, and P. Butaud,
\newblock Interaction induced delocalization for two particles in a periodic potential,
\newblock \textit{Phys. Rev. Lett.} \textbf{85}, 3906 (2000).


\bibitem{vidal2001disorder}
J. Vidal, P. Butaud, B. Dou\c{c}ot, and R. Mosseri,
\newblock Disorder and interactions in Aharonov-Bohm cages,
\newblock \textit{Phys. Rev. B} \textbf{64}, 155306 (2001).

\bibitem{Tovmasyan2013}
M. Tovmasyan, E. P. L. van Nieuwenburg, and S. D. Huber,
\newblock Geometry-induced pair condensation,
\newblock \textit{Phys. Rev. B} \textbf{88}, 220510(R) (2013).


\bibitem{Tovmasyan2018}
M. Tovmasyan, S. Peotta, L. Liang, P. T\"orm\"a, and S. D. Huber,
\newblock Preformed pairs in flat Bloch bands,
\newblock \textit{Phys. Rev. B} \textbf{98}, 134513 (2018).


\bibitem{cartwright2018rhombi}
C. Cartwright, G. De Chiara, and M. Rizzi,
\newblock Rhombi-chain Bose-Hubbard model: Geometric frustration and interactions,
\newblock \textit{Phys. Rev. B} \textbf{98}, 184508 (2018).

\bibitem{liberto2019nonlinear}
M. Di Liberto, S. Mukherjee, and N. Goldman,
\newblock Nonlinear dynamics of Aharonov-Bohm cages,
\newblock \textit{Phys. Rev. A} \textbf{100}, 043829 (2019).


\bibitem{danieli2021quantum}
C. Danieli, A. Andreanov, T. Mithun, and S. Flach,
\newblock Quantum caging in interacting many-body all-bands-flat lattices,
\newblock \textit{Phys. Rev. B} \textbf{104}, 085132 (2021).

\bibitem{danieli2021nonlinear}
C. Danieli, A. Andreanov, T. Mithun, and S. Flach,
\newblock Nonlinear caging in all-bands-flat lattices,
\newblock \textit{Phys. Rev. B} \textbf{104}, 085131 (2021).


\bibitem{tovmasyan2018preformed}
M. Tovmasyan, S. Peotta, L. Liang, P. T\"orm\"a, and S. D. Huber,
\newblock Preformed pairs in flat Bloch bands,
\newblock \textit{Phys. Rev. B} \textbf{98}, 134513 (2018).

\bibitem{martinez2023flat}
J. G. C. Martinez, C. S. Chiu, B. M. Smitham, and A. A. Houck,
\newblock Flat-band localization and interaction-induced delocalization of photons,
\newblock \textit{Sci. Adv.} \textbf{9}, eadj7195 (2023).

\bibitem{chen2025interaction}
T. Chen, C. Huang, I. Velkovsky, T. Ozawa, H. Price, J. P. Covey, and B. Gadway,
\newblock Interaction-driven breakdown of Aharonov--Bohm caging in flat-band Rydberg lattices,
\newblock \textit{Nat. Phys.} \textbf{21}, 221 (2025).

\bibitem{tai2017microscopy}
M. E. Tai, A. Lukin, M. Rispoli, R. Schittko, T. Menke, D. Borgnia, P. M. Preiss, F. Grusdt, A. M. Kaufman, and M. Greiner,
\newblock Microscopy of the interacting Harper--Hofstadter model in the two-body limit,
\newblock \textit{Nature} \textbf{546}, 519 (2017).

\bibitem{pan2025reversal}
L. Pan, Q. Liang, C.-A. Yang, Y. Huang, P. Liu, F. Xi, W. Yi, X. Zhou, and J.-S. Pan,
\newblock Interaction-induced chiral-transport inversion,
\newblock \textit{Phys. Rev. A} \textbf{112}, L061303 (2025).


\bibitem{greschner2015spontaneous}
S. Greschner, M. Piraud, F. Heidrich-Meisner, I. P. McCulloch, U. Schollw\"ock, and T. Vekua,
\newblock Spontaneous Increase of Magnetic Flux and Chiral-Current Reversal in Bosonic Ladders: Swimming against the Tide,
\newblock \textit{Phys. Rev. Lett.} \textbf{115}, 190402 (2015).

\bibitem{kolley2015strongly}
F. Kolley, M. Piraud, I. P. McCulloch, U. Schollw\"ock, and F. Heidrich-Meisner,
\newblock Strongly interacting bosons on a three-leg ladder in the presence of a homogeneous flux,
\newblock \textit{New J. Phys.} \textbf{17}, 092001 (2015).

\bibitem{uchino2016analytical}
S. Uchino,
\newblock Analytical approach to a bosonic ladder subject to a magnetic field,
\newblock \textit{Phys. Rev. A} \textbf{93}, 053629 (2016).

\bibitem{Wang2008}
Z.-X. Wang,
\newblock \textit{Concise Quantum Field Theory},
\newblock (Peking University Press, Beijing, 2008).

\bibitem{Zhao2025}
H. Zhao, H. Du, Y. Wang, Y. Li, C. Ren, J. Wu, W. Liu, L. Xiao, S. Jia, Q. Fan, and J. Ma,
\newblock Observation of interaction-induced asymmetric chiral currents in a synthetic zigzag ladder,
\newblock \textit{Phys. Rev. A} \textbf{112}, 023307 (2025).


\end{thebibliography}
\end{document}